\documentclass{birkmult}
\usepackage{graphicx}        

\usepackage {latexsym}
\usepackage{natbib}

 \newtheorem{thm}{Theorem}[section]

 \newtheorem{prop}[thm]{Proposition}
 \theoremstyle{definition}
 
 \theoremstyle{remark}

 \numberwithin{equation}{section}
\graphicspath{%
    {converted_graphics/}
    {./}
}
\begin{document}

\title{Chaos in the St\"{o}rmer problem}

\author{Rui Dil\~ao} 

\address{Nonlinear Dynamics Group, Instituto Superior T\'ecnico\\
Av. Rovisco Pais, 1049-001 Lisbon, Portugal}

\email{rui@sd.ist.utl.pt; ruidilao@gmail.com}

\author{Rui Alves-Pires}

\address{Faculdade de Engenharia, Universidade Cat\' {o}lica Portuguesa\\
Estrada de Tala\'{i}de, 2635-631 Rio de Mouro, Portugal}

\email{pires@fe.ucp.pt; pires@sd.ist.utl.pt}

\begin{abstract}
We survey the few exact results on the St\"{o}rmer problem describing
the dynamics of charged particles in the Earth magnetosphere. 
The analysis of this system leads to the the conclusion that charged particles are trapped in the Earth magnetosphere or escape to infinity, and the trapping region is bounded by a torus-like surface, the Van Allen inner radiation belt. In the trapping region, the motion of the charged particles can be periodic, quasi-period or chaotic. The three main effects observed in the Earth magnetosphere, radiation belts, radiation aurorae and South Atlantic anomaly,  
are described in the framework  described here.
We discuss some new mathematical problems suggested 
by the analysis of the St\"{o}rmer problem.
\end{abstract}




\maketitle

\section{Introduction}\label{s1}

Stellar and planetary magnetic environments or magnetospheres
are generated by the motion of charged particles inside the core of stars and planets.
In the magnetosphere of the Earth, incoming charged particles have intricate trajectories
and are in the origin of observable radiation
phenomena as is the case of radiation aurorae \citep{stormer}, the Van Allen inner radiation belts \citep{VanAllen},
and the South Atlantic anomaly \citep{Dilao}.

The magnetic field of the Earth has a strong dipolar component, \citep{Rikitake}, and it is believed that, for low altitudes ($\le 3000\, km$), the radiation phenomena occurring in the Earth magnetosphere can be understood by studying the motion of nonrelativistic charged particles in a pure dipole field. If the dipolar component of the Earth magnetic field    is considered aligned with the rotation axis of the Earth, the equations of motion of a charged particle in the  dipolar field of the  Earth reduce to a non-linear autonomous  Hamiltonian dynamical system. This is the St\"{o}rmer problem, \citep{stormer}.

The analysis of the St\"{o}rmer  problem presents big challenges from the theoretical, applied and computational points of view. 

From the computational point of view, the determination  of the
trajectories of high energy charged particles in the Earth magnetosphere for long periods of time is   inaccurate and time consuming, being difficult to extract information about the several aspects of the radiation phenomena observed in the Earth magnetosphere. For example, in order to explain some of the dynamic aspects associated with aurorae and magnetic mirrors, 
adiabatic {\it ad-hoc} arguments have been introduced 
into the theory, and the motion of charged particles in the magnetosphere has been assumed similar to the cyclotron type motion  
in constant magnetic fields, 
\citep{VanAllen}, \citep{hess} and \citep{daly}. 

The long lived and transient radiation belts
observed 
in the Earth magnetosphere  have adverse effects on the electronics of spacecrafts, and affect communications, \citep{stassinopoulos} and \citep{daly}. 
Therefore, a qualitative and quantitative understanding of the 
St\"{o}rmer problem has important applications.

The first theoretical studies about the properties of the trajectories of  charged particle in a dipolar field where done by   \cite{DeVogelaere} and \cite{Dragt}. In the work of these authors,
the existence of a trapping region for charged particles in the dipole field of the Earth was implicitly established.

Based on the qualitative theory of conservative maps of the plane, 
\cite{DragtFinn} carried a comparative study between the phase space topology of the orbits of a generic area-preserving 
map of the plane and the numerically computed Poincar\'{e} sections 
of the St\"{o}rmer problem. They have presented numerical evidence about 
the existence of homoclinic points in the Poincar\'{e} sections. According to these authors, this
shows that the St\"{o}rmer problem is insoluble, implying that the adiabatic magnetic moment series diverges, a basic theoretical argument used by Van Allen to explain some features of the radiation phenomena in planetary magnetospheres.
The KAM approach to the St\"{o}rmer  problem has been developed by Braun in a sequence of papers, \citep{Braun1}, \citep{Braun2} and \citep{Braun4}. Within this approach, it has been shown that trapped particles can have quasi-periodic    motion, and can penetrate arbitrarily close to the dipole axis. 

Due to its intrinsic difficulty, the analysis of the St\"{o}rmer problem
has been done using a mixture of analytical and numerical techniques. 
Here, we are interested in surveying the exact results on the 
St\"{o}rmer problem,  separating the
results that are numeric from the exact ones. All the exact results are summarized in Propositions \ref{P1} and \ref{P2}. We have made extensive simulations of trajectories of charged particles in the Earth 
magnetosphere, and we have obtained the shape of the trapping regions for charged particles --- Van Allen inner radiation belts. This contrast with the usual approach used in radiation environment studies, where
radiation belt boundaries are correlated with the dipole field lines, \citep{daly}.

This paper is organized as follows. In the next section, we derive the equations of motion for the St\"{o}rmer problem and we obtain its conservation laws. In section \ref{s4}, we study the motion of charged particles 
in the equatorial plane of the Earth. In section \ref{s5}, we analyse the general case for the motion on the three-dimensional configuration space. In the final section \ref{s6}, we summarize the main results
from the theoretical and applied points of views,
and we discuss some of the mathematical problems suggested 
by the analysis done previously.

\section{Equations of motion and  conservation laws}\label{s2}

The equation of motion of a nonrelativistic charged particle of mass 
$m$ and  charge $q$ in a magnetic field $\vec {B}$ has the Lorentz form, 
\begin{equation}
m\vec {\ddot {r}} = q\left( {\vec {\dot {r}}\times \vec {B}} \right)\, .
\label{eq1}
\end{equation}
We use the international system of units and 
$B$ is measured in Tesla. At the surface of the Earth, $B$ is in the range $(0.5\times 10^{-4}-1.0\times 10^{-4})$ Tesla $= (0.5-1.0)$ Gauss.

Magnetic fields are produced by moving charges and currents. 
A magnetic dipole field can be produced by a current loop on a planar surface, and the resulting field is proportional to the current intensity times the area delimited by the current loop, (\citep{Fey}, pp. 14-7). 
For a current loop in the horizontal $xy$-plane, flowing counterclockwise with current
intensity $I$,  the dipole momentum is 
$\vec{\mu}_z=\mu \vec{e}_z$, where 
$\mu=I\times area\ of\ the \ loop$, and $\vec{e}_z$ is the unit vector  of the $z$ axis. This current loop 
produces a dipole field with a dipole momentum pointing in the positive direction of the 
$z$-axis. This dipole field derives from the vector 
potential,  
\begin{equation}\displaystyle
\vec{A}={1\over 4 \pi \varepsilon_0 c^2}{1\over r^2}\  \vec{\mu}_z\times \vec{e}_r ={1\over 4 \pi \varepsilon_0 c^2}{1\over r^3}\ \vec{\mu}_z\times \vec{r} =M_z {1\over r^3} \ (-y\vec{e}_x+x\vec{e}_y)
\label{eq2}
\end{equation}
where $r=\sqrt{x^2+y^2+z^2}$, $M_z$ is the (scalar) dipole momentum, 
for short, and the 
vector potential is independent of the $z$ coordinate. As $\vec{B}=\hbox{rot}  \vec{A}$, by  (\ref{eq2}), the Lorentz equation 
(\ref{eq1}) describing the motion of a charge particle in a dipole field is,
\begin{equation} 
\left\{
\begin{array}{l}\displaystyle
\ddot{x}=3\alpha {z\over r^5}  (\dot y z-\dot z y)-\alpha \dot y {1\over r^3}\\ [8pt]\displaystyle
\ddot{y}=-3\alpha {z\over r^5}  (\dot x z-\dot z x)+\alpha \dot x {1\over r^3}\\ [8pt]\displaystyle
\ddot{z}=3\alpha{z\over r^5}  (\dot x y-\dot y x)
\end{array}\right.
\label{eq3}
\end{equation}
where $\alpha=qM_z/m$. For the Earth dipolar field, the dipole momentum is  
$M_z= 7.9\times 10^{25}\ \hbox{G\, cm}^3=7.9\times 10^{15}\ \hbox{T\, m}^3$ (\citep{Rikitake}, 1975 IGRF value), and, for electrons and protons, we have,
\begin{equation}
\alpha=\left\{
\begin{array}{l}
-1.45\times 10^{27} \ \hbox{m}^3/\hbox{s}\ (\hbox{electrons})\\ [5pt]
7.88\times 10^{23} \ \hbox{m}^3/\hbox{s}\ (\hbox{protons})\, .
\end{array}\right.
\label{eq4}
\end{equation}

We now rescale the system of equations (\ref{eq3}). With 
$X=x/r_0$, $Y=y/r_0$, and $Z=z/r_0$, where
$r_0=6378136\ \hbox{m}$ is the radius of the Earth, we rewrite  the
system of equations (\ref{eq3}) in the form,
\begin{equation}
\left\{
\begin{array}{l}\displaystyle
\ddot{X}=3\alpha_1 {Z\over R^5}  (\dot Y Z-\dot Z Y)-\alpha_1 \dot Y {1\over R^3}\\ [8pt]\displaystyle
\ddot{Y}=-3\alpha_1 {Z\over R^5}  (\dot X Z-\dot Z X)+\alpha_1 \dot X {1\over R^3}\\ [8pt]\displaystyle
\ddot{Z}=3\alpha_1 {Z\over R^5}  (\dot X Y-\dot Y X)
\end{array}\right.
\label{eq9}
\end{equation}
where $R=\sqrt{X^2+Y^2+Z^2}$, 
\begin{equation}\displaystyle
\alpha_1={\alpha\over r_0^3}=\left\{
\begin{array}{l}
-5.588\times 10^{6} \ \hbox{s}^{-1}\ (\hbox{electrons})\\ [5pt]
3.037\times 10^{3} \  \hbox{s}^{-1}\ (\hbox{protons})\, .
\end{array}\right.
\label{eq10}
\end{equation}
In this rescaled coordinate system, if $R(t)=1$, the charged particle hits the surface of the Earth. 

The Lorentz equations (\ref{eq9}) can be derived from a Lagrangian. By standard Lagrangian mechanics techniques, and by (\ref{eq2}), we have,
\begin{equation}
\begin{array}{lcl} 
L&=&\displaystyle {1\over 2} m ({\dot x}^2+{\dot y}^2+{\dot z}^2)+q \vec{\dot r}.\vec{A}= {1\over 2} m ({\dot x}^2+{\dot y}^2+{\dot z}^2)
-{m\alpha\over r^3}(\dot x y-\dot y x)
\\[8pt] 
&=&\displaystyle {1\over 2} m ({\dot X}^2+{\dot Y}^2+{\dot Z}^2)
-{m\alpha_1\over R^3}(\dot X Y-\dot Y X)\, .
\end{array} 
\label{eq12}
\end{equation}
The conjugate momenta to the coordinates $X$, $Y$ and $Z$ are,
\begin{equation}
\begin{array}{l}\displaystyle
p_X={\partial L\over \partial \dot X}=m\dot X-m\alpha_1 {Y\over R^3}\\ [8pt]\displaystyle
p_Y={\partial L\over \partial \dot Y}=m\dot Y+m\alpha_1 {X\over R^3}\\ [8pt]\displaystyle
p_Z= {\partial L\over \partial \dot Z}=m\dot Z
\end{array}
\label{eq6}
\end{equation}
and  the Hamiltonian is,
\begin{equation}
\begin{array}{lcl}
H&=&\displaystyle \vec{p}.\dot{\vec R}-L={1\over 2} m ({\dot X}^2+{\dot Y}^2+{\dot Z}^2)
\\[8pt] 
&=& \displaystyle {1\over 2m} ({p_X}^2+{p_Y}^2+{p_Z}^2)-\alpha_1 p_Y {X\over R^3}+
\alpha_1 p_X {Y\over R^3}+{m\alpha_1^2\over 2}{X^2+Y^2\over R^6}\, .
\end{array}
\label{eq7}
\end{equation}
Hence, the system of equations (\ref{eq9}) has the conservation law,
\begin{equation}
H={1\over 2} m ({\dot X}^2+{\dot Y}^2+{\dot Z}^2)\, .
\label{eq8}
\end{equation}

We show now that the  system of equations  (\ref{eq9}) has a second constant of motion.

To determine the second constant of motion, we introduce cylindrical coordinates. With, $X=\rho \cos\phi$, and $Y=\rho \sin\phi$,
the Lagrangian (\ref{eq12}) becomes,
\begin{equation}
L'= \displaystyle {1\over 2} m ({\dot \rho}^2+{\rho}^2 {\dot \phi}^2+{\dot Z}^2)
+{m\alpha_1\over R^3}{\rho}^2 {\dot \phi} 
\label{eq13}
\end{equation}
where $R=\sqrt{\rho^2+Z^2}$. In this coordinate system, the conjugate momenta become,
\begin{equation}
\begin{array}{l}\displaystyle
p_{\rho}={\partial L\over \partial \dot \rho}=m\dot \rho\\ [8pt]\displaystyle
p_{\phi}={\partial L\over \partial \dot \phi}=m{\rho}^2 {\dot \phi} +m\alpha_1 \rho^2{1\over R^3}\\ [8pt]\displaystyle
p_Z= {\partial L\over \partial \dot Z}=m\dot Z
\end{array}
\label{eq14}
\end{equation}
and the new Hamiltonian is now,
\begin{equation}\displaystyle 
H={1\over 2} m ({\dot X}^2+{\dot Y}^2+{\dot Z}^2)
= {1\over 2m} \left({p_{\rho}}^2+{p_Z}^2+\left({p_{\phi}\over \rho}-m\alpha_1 {\rho\over R^3}\right)^2 \right)\, .
\label{eq15}
\end{equation}
As the Hamiltonian (\ref{eq15}) is independent of $\phi$, we have 
$\displaystyle \dot{p}_{\phi}=-{\partial H\over \partial \phi}=0$, and by (\ref{eq14}), the second conservation law is,
\begin{equation}\displaystyle
p_{\phi}={\partial L\over \partial \dot \phi}=m{\rho}^2 {\dot \phi} +m\alpha_1 \rho^2{1\over R^3}=c_2=constant \, .
\label{eq16}
\end{equation}
As $p_{\phi}=c_2$ is the second equation of motion, we can write the Hamiltonian (\ref{eq15}) in the form,
\begin{equation}\displaystyle
H
= {1\over 2m} \left({p_{\rho}}^2+{p_Z}^2+\left({c_2\over \rho}-m\alpha_1 {\rho\over R^3}\right)^2 \right)\, .
\label{eq17}
\end{equation}
Then, the equations of motion of a nonrelativistic charged particle in a dipole field reduce to,
\[
\left\{
\begin{array}{l}\displaystyle
\dot \rho = {\partial H\over \partial p_{\rho}}={1\over m} p_{\rho}\\[8pt]\displaystyle
\dot p_{\rho} = -{\partial H\over \partial  \rho}={1\over m\rho^3} c_{2}^2+3m \alpha_1^2{\rho^3\over R^8}-m\alpha_1^2{\rho \over R^6}
-3\alpha_1c_2{\rho \over R^5}\\[8pt]\displaystyle
\dot Z = {\partial H\over \partial p_{Z}}={1\over m} p_{Z}\\[5pt]
\dot p_{Z} = -{\partial H\over \partial  Z}=3m \alpha_1^2{\rho^2 Z\over R^8}
-3\alpha_1c_2{Z \over R^5}
\end{array}\right.
\]
or,
\begin{equation}
\left\{
\begin{array}{l}\displaystyle
\ddot \rho ={1\over m^2\rho^3} c_{2}^2+3  \alpha_1^2{\rho^3\over R^8}- \alpha_1^2{\rho \over R^6}
-3{\alpha_1c_2\over m}{\rho \over R^5}\\[8pt]\displaystyle
\ddot Z = 3 \alpha_1^2{\rho^2 Z\over R^8}
-3{\alpha_1c_2\over m}{Z \over R^5}
\end{array}\right.
\label{eq18}
\end{equation}
together with the conservation laws,
\begin{equation}
\left\{
\begin{array}{l}\displaystyle
{1\over 2m^2} \left({p_{\rho}}^2+{p_Z}^2+\left({c_2\over \rho}-m\alpha_1 {\rho\over R^3}\right)^2 \right)=c_1\\[8pt]\displaystyle
m{\rho}^2 {\dot \phi} +m\alpha_1 \rho^2{1\over R^3}=c_2 
\end{array}\right.
\label{eq19}
\end{equation}
where $R=\sqrt{\rho^2+Z^2}$. Note that, we have rescaled the Hamiltonian of the St\"{o}rmer problem to, 
\begin{equation}
H_{eff}=
{1\over 2} \left({\dot \rho}^2+{\dot Z}^2\right)+{1\over 2 m^2}\left({c_2\over \rho}-m\alpha_1 {\rho\over R^3}\right)^2  =T+V_{eff}(\rho ,Z)
\label{eq20}
\end{equation}
where $T$ is a scaled kinetic energy.

Hence, the  motion of charged particle in a dipole field is described by a two-degree of freedom Hamiltonian system. The effective Hamiltonian is parameterized by  the second conservation law in (\ref{eq19}), which determines the time dependence of the angular cylindrical coordinate.
 
Before analysing the general topology of the orbits of the phase space flow of equations (\ref{eq18}), we first consider the case where the motion restricted to the plane $Z=0$.

\section{Motion in the equatorial plane of the dipole field}\label{s4}

Here, we consider the   case where charged particles are constrained to the
equatorial plane of the Earth, the 
plane $Z=0$, for every $t\ge 0$. In this case, by (\ref{eq18}), the equations of motion reduce to,
\begin{equation}\displaystyle
\ddot \rho ={c_{20}^2 \over m^2 } {1\over  \rho^3} 
-3{\alpha_1c_{20}\over m}{1 \over \rho^4}+2  \alpha_1^2{1\over \rho^5}=-
{d{\bar V}_{eff}\over d\rho}
\label{se4eq1}
\end{equation}
where $c_{20}$ is the value of the constant $c_2$ evaluated at $Z=0$, and, by (\ref{eq20}), the potential function $V_{eff}(\rho)$ is,
\begin{equation}\displaystyle
{\bar V}_{eff}(\rho)={c_{20}^2 \over 2 m^2 } {1\over  \rho^2} 
-{\alpha_1c_{20} \over m}{1 \over \rho^3}+{\alpha_1^2\over 2}{1\over \rho^4} \, .
\label{se4eq2}
\end{equation}
Introducing $p_Z=0$ and $Z=0$ into (\ref{eq19}), the two conservation laws reduce to,
\begin{equation}
\begin{array}{l}\displaystyle
{1\over 2m^2} \left({p_{\rho}}^2 +\left({c_{20}\over \rho}-m\alpha_1 {1\over \rho^2}\right)^2 \right)=c_{10}\\[8pt]\displaystyle
m{\rho}^2 {\dot \phi} +m\alpha_1  {1\over \rho}=c_{20} 
\end{array}
\label{se4eq3}
\end{equation}
where,
\begin{equation}
\left\{
\begin{array}{l}\displaystyle
c_{10}={1\over 2} {\dot \rho (0)}^2 +{1\over 2m^2}\left({c_{20}\over \rho(0)}-m\alpha_1 {1\over \rho^2(0)}\right)^2  \\[8pt]\displaystyle
c_{20}=m{\rho(0)}^2 {\dot \phi(0)} +m\alpha_1  {1\over \rho(0)} 
\end{array}\right.
\label{se4eq4}
\end{equation}
and $c_{10}$ is the value of the effective total energy   evaluated at $Z=0$.

Integrating  the second conservation law in (\ref{se4eq3}) by quadratures, we obtain, for the angular coordinate $\phi(t)$,
\begin{equation}\displaystyle
\phi(t)=\phi(0)+\int_0^t \left( {c_{20}\over m \rho(s)^2}-{\alpha_1\over \rho(s)^3}
\right) dt\, .
\label{se4eq5}
\end{equation}
So, if the solution $\rho(t)$ of equation (\ref{se4eq1}) is known, the temporal dependency of angular coordinate is obtained from (\ref{se4eq5}).

\begin{figure}
\centering
\includegraphics[width=11.5cm]{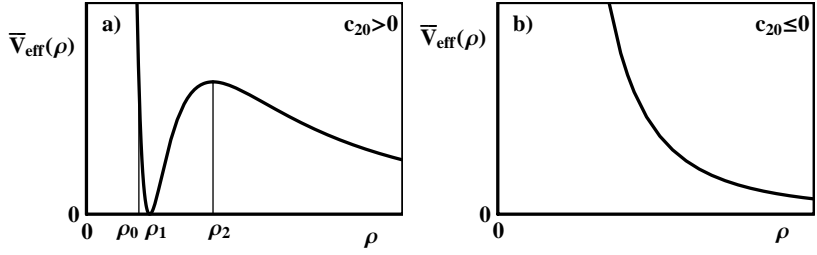}
\caption{Effective potential associated with the motion of a charged particle (protons, $\alpha_1 >0$) in the equatorial plane of the Earth ($Z=0$). a) If $c_{20}>0$ and $\rho\ge 0$, the effective potential has a maximum for $\rho=\rho_2$, a minimum for $\rho=\rho_1$, ${\bar V}_{eff}(\rho_1)=0$, and ${\bar V}_{eff}(\rho_2)={c_{20}^4/ (32m^4 \alpha_1^2)}$. At $\rho=\rho_0= (\sqrt{2}-1)\rho_2$, ${\bar V}_{eff}(\rho_0)={\bar V}_{eff}(\rho_2)$. b) If $c_{20}\le 0$ and $\rho\ge 0$, ${\bar V}_{eff}(\rho)$ is a monotonically decreasing function of the argument, and the effective energy surfaces have no compact components.}
\label{fig1}
\end{figure}

In these conditions, the motion of the charged particle in the equatorial plane of the Earth is completely determined by equation (\ref{se4eq1}), derived from the effective Hamiltonian, 
\begin{equation} 
{\bar H}_{eff}(\rho,\dot \rho)={\bar T}+{\bar V}_{eff}=\displaystyle {1\over 2}\dot \rho^2+{c_{20}^2 \over 2 m^2 } {1\over  \rho^2} 
-{\alpha_1c_{20}\over m}{1 \over \rho^3}+{\alpha_1^2\over 2}{1\over \rho^4}
\label{se4eq7}
\end{equation}
where ${\bar T}$ is a scaled kinetic energy.
If $c_{20}>0$  and for $\rho >0$, the potential function ${\bar V}_{eff}(\rho)$
has one local minimum and one local maximum at,
\begin{equation}
\rho_1={m\alpha_1\over c_{20}}\quad \hbox{and}\quad \rho_2=2{m\alpha_1\over c_{20}}
\label{se4eq6}
\end{equation}  
respectively, Figure \ref{fig1}a).
By direct calculation, we have, ${\bar V}_{eff}(\rho_2) = {c_{20}^4/ (32m^4 \alpha_1^2)}$, and ${\bar V}_{eff}(\rho_1) = 0$. Therefore, 
for $0\le {\bar H}_{eff}(\rho,\dot \rho)\le {\bar V}_{eff}(\rho_2)$,
the constant effective  energy surface contains a compact component.

If $c_{20}\le 0$  and for positive values of $\rho$, the potential function ${\bar V}_{eff}(\rho)$ is a
monotonically decreasing function of the argument, Figure \ref{fig1}b).  Then, we have:

\begin{prop}\label{P1} In the equatorial plane of a dipole field, a nonrelativistic (positively) charged particle precesses around the dipole axis, along a circle with radius $\rho_1$, provided:
(i) ${\bar H}_{eff}(\rho(0),\dot \rho(0))<{\bar V}_{eff}(\rho_2)$, $c_{20}>0$, $\rho(0)\not=\rho_1$ and $\rho(0)<\rho_2$, where ${\bar H}_{eff}$ is defined in (\ref{se4eq7}),  the constants $\rho_1$ and $\rho_2$   are given in (\ref{se4eq6}), and the constant $c_{20}$ is defined in (\ref{se4eq4}).  For initial conditions near the circumference of radius $\rho=\rho_1$,  the Larmor or precession period is, $T_L=2\pi m^3\alpha_1^2/c_{20}^3$. The phase advance per Larmor period is,
\[\displaystyle
\Delta \phi = 2\pi  \frac{\rho_1^2}{\rho(0)^2}\left(1-\frac{\rho_1}{\rho(0)}\right)-
4\pi^2 \frac{\rho_1^3}{\rho(0)^3}\frac{m}{c_{20}}\rho_1 {\dot\rho}(0)\left(1-\frac{3}{2}\frac{\rho_1}{\rho(0)}\right)+\cdots
\]
and the period of rotation around the Earth is $T_r=2\pi T_L/\Delta \phi$.

(ii) If ${\bar H}_{eff}(\rho(0),\dot \rho(0))={\bar V}_{eff}(\rho_2)$, $c_{20} >0$, and
$\rho(0)<\rho_2$, or, $\rho(0)>\rho_2$ and ${\dot \rho}(0)<0$,  then, $\rho(t)\to \rho_2$, as
$t \to \infty$. 

(iii) If $\rho(0)=\rho_2$ [resp., $\rho(0)=\rho_1$], $\dot \rho(0)=0$, 
${\dot\phi}(0)\not=0$, and $c_{20} >0$,
then the charged particle has a circular trajectory with radius $\rho=\rho_2$ [resp., $\rho=\rho_1$], and the period of rotation around the Earth is $T_r=2\pi/\dot \phi(0)$. If
${\dot\phi}(0)=0$, the charged particle is at rest.

(iv) If ${\bar H}_{eff}(\rho(0),\dot \rho(0))>{\bar V}_{eff}(\rho_2)$ and
$c_{20}>0$, or, $c_{20}\le 0$, then, $\rho(t)\to \infty$, as
$t \to \infty$, and we have escape trajectories.
\end{prop}

\begin{proof} To prove the proposition, we must ensure first that the solution of the differential equation (\ref{se4eq1}) exists, and 
is defined for every $t\ge 0$. In the cases {\it(i)}-{\it(iii)},
the existence of solutions for  every $t\ge 0$ follows because 
the initial conditions on phase space are on the compact components 
of the level sets of the Hamiltonian function (\ref{se4eq7}), 
(\cite{chi}, pp. 187; \cite{nemitskii}, pp. 8). 

A simple phase space analysis shows that the vector field associated with equation (\ref{se4eq1}) has 
a centre type fixed point with phase space (cylindrical) coordinates $(\rho_1,{\dot \rho}=0)$. This centre type fixed point is inside the homoclinic loop of a saddle point with coordinates $(\rho_2,{\dot \rho}=0)$. The conditions in {\it(i)} correspond  to initial conditions inside the homoclinic loop, and away
from the centre fixed point,  $\rho(0)\not=\rho_1$. 
To calculate the  Larmor frequency and the phase advance per Larmor period, we linearize equation (\ref{se4eq1}) around 
$\rho_1$, and we obtain,
\[\displaystyle
\ddot x+\frac{c_{20}^6}{m^6 \alpha_1^4}x=0
\]
where $x=\rho-\rho_1$, and ${c_{20}^6}/{(m^6 \alpha_1^4)}=\omega_L^2$. 
The Larmor period is $T_L=2\pi/ \omega_L$. Note that, by a direct calculation,  
$\omega_L^2=\displaystyle {d^2{\bar V}_{eff}\over d\rho^2}(\rho=\rho_1)$.

By (\ref{se4eq5}), the phase advance per Larmor period is,
\[ \displaystyle
\Delta \phi = \int_0^{T_L} \left( {c_{20}\over m \rho(s)^2}-{\alpha_1\over \rho(s)^3}
\right) dt
\]
From the above linearized differential equation, in the vicinity of the circumference of radius $\rho=\rho_1$,  $\rho(t)=\rho_1+(\rho-\rho_1)\cos(\omega_Lt)+{\dot \rho}(0)\sin(\omega_Lt)/\omega_L$ and after substitution into the
expression for $\Delta \phi$, we obtain the result. 
The  rotation
period around the Earth is obtained from the conditions, $n\Delta \phi=2\pi $  and $T_r=nT_L$.

The conditions in {\it(ii)} correspond to initial conditions on the stable branches of the homoclinic orbit of the saddle point with
coordinates $(\rho_2,{\dot \rho}=0)$.

For case {\it (iii)} , as $(\rho_2,{\dot \rho}=0)$ is a fixed point of the differential equation (\ref{se4eq1}), if $\dot\phi(0)\not=0$, the charged particle has a circular trajectory around the origin of coordinates. 
To calculate the period of the trajectory, by (\ref{se4eq5}), we have,
\[\displaystyle
2\pi = \left( {c_{20}\over m \rho_2^2}-{\alpha_1\over \rho_2^3}
\right) T_r
\]
Introducing the value of the constant $c_{20}$ into the above expression, by (\ref{se4eq4}), we obtain $T_r=2\pi/\dot \phi(0)$. For $\rho(0)=\rho_1$, the proof is similar. 

In case {\it (iv)}, the effective potential function  monotonically decreases as $\rho$ increases from $\rho=0$, and we are in the conditions of escape trajectories. The prolongation of solutions for
every $t\ge 0$, follows from the condition that $f_i(\rho)/||\rho||\to constant$ as $\rho \to \infty$ (\cite{nemitskii}, pp. 9).
\end{proof}

\bigskip

\begin{figure}
\centering
\includegraphics[width=9.0cm]{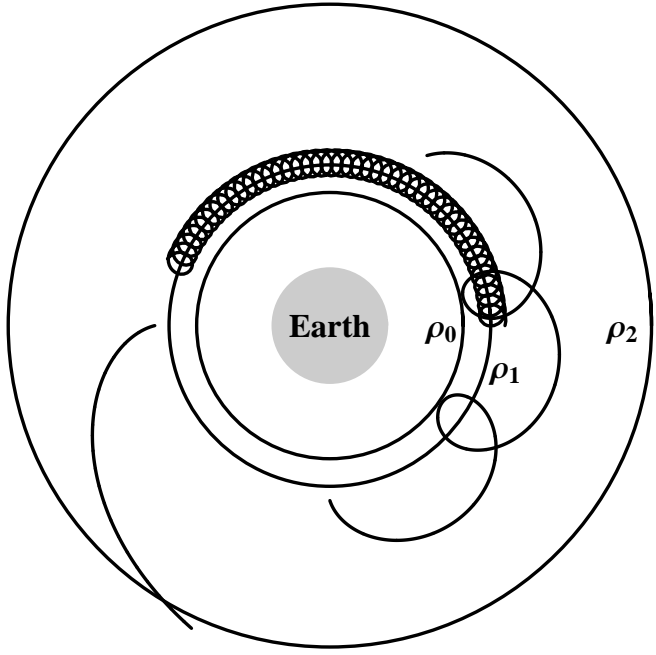}
\caption{Equatorial cross section of the Van Allen inner radiation belt, for protons in the equatorial plane of the Earth. We show the trajectories of three protons with initial conditions (cylindrical coordinates): a) $\rho (0)=3.0$,
$\dot \rho(0)=10.0$, $\phi(0)=0.0$ and $\dot \phi (0)=10.0$; b) $\rho (0)=3.0$,
$\dot \rho(0)=80.0$, $\phi(0)=3\pi /2$ and $\dot \phi (0)=10.0$; c) $\rho (0)=3.0$,
$\dot \rho(0)=100.0$, $\phi(0)=\pi $ and $\dot \phi (0)=10.0$. The first and the second trajectories, a) and b), correspond to the precession of  protons around the Earth. The third case is an escape trajectory. By (\ref{se4eq8}), the escape condition is  $|\dot \rho(0)|\ge 95.42$. For the three trajectories shown, the radiation belt parameters are $\rho_0=2.282$, $\rho_1=2.755$,
$\rho_2=5.510$, and $c_{20}=1.844\times 10^{-24}$. The total effective energies are: a)
$c_{10}=500$; b)  $c_{10}=3650$, and,    c)  $c_{10}=5450$.  The trajectories have been calculate with the 
St\"{o}rmer-Verlet  numerical method (Appendix) with the time step $\Delta t=0.0001$, and the angular coordinate has been calculated by (\ref{se4eq9}).  
For protons with precessing trajectories near the circumference of radius $\rho_1$, by Proposition \ref{P1}, the Larmor period is $T_L=0.043\, $s.  The first two particle trajectories have been calculated from $t=0$ up to $t=2\, $s.}
\label{fig2}
\end{figure}

By Proposition \ref{P1}, the trajectory of
a charged particle on the equatorial plane of an axially symmetric magnetosphere can be unbounded, can be trapped in an annular region around the Earth, or
can collide with the surface of the Earth. As $\rho_0$, $\rho_1$ and
$\rho_2$ depend
on $c_{20}$, there are charged particles with  different energies trapped in the annular region $[\rho_0,\rho_2]$.
This annular region is the equatorial cross section of the Van Allen inner radiation belt, \citep{VanAllen}. 

As $\rho_0$, $\rho_1$ and $\rho_2$ depend on the initial conditions
$\rho(0)$ and $\dot \phi(0)$ through $c_{20}$, for the same belt parameter, there are particles that escape from the trapping region, escaping to infinity or hitting 
the surface of the Earth. 
This is due to the fact that the belt parameters are
independent of $\dot \rho(0)$. By Proposition \ref{P1}-{\it (iv)}, and by the conservation of energy, a charged particle   escapes  from the inner Van Allen radiation belt if, 
\begin{equation}
|\dot \rho (0)|> \sqrt{2|{\bar V}_{eff}(\rho_2)-{\bar V}_{eff}(\rho(0))|}
\label{se4eq8}
\end{equation}
By the same argument, if  $|\dot \rho (0)|\ge |c_{20}/m -\alpha_1|$,
a particle with initial velocity $\dot \rho (0)<0$ at infinity, and velocity vector within the equatorial plane of the Earth, after an infinite time, hits the surface of the Earth.

 In Figure \ref{fig2}, we show the limits of the equatorial cross section of a 
Van Allen inner radiation belt, the trajectories of trapped protons, and
an escape trajectory. In the three cases shown, the Van Allen parameters are the same. The  trajectories shown in Figure \ref{fig2} have been calculated by the numerical integration of equation (\ref{se4eq1}), (see the Appendix). The angular variable $\phi $ has been obtained from
the discretization of (\ref{se4eq5}),
\begin{equation}
\phi_{n+1}=\phi_{n}+\Delta t \left( {c_{20}\over m \rho_n^2}-{\alpha_1\over \rho_n^3} \right)  
\label{se4eq9}
\end{equation}
where $\Delta t$ is the discretization time step, $\phi_n=\phi(n \Delta t)$,  $\rho_n=\rho(n \Delta t)$, and $n=0,1,\cdots$.

As the radiation  belt parameters 
$\rho_0$, $\rho_1$ and $\rho_2$  depend  on the angular velocity $\dot \phi(0)$, we can have several Van Allen radiation belts  at different altitudes, and trapped particles with different energies.  

\section{Motion in the three-dimensional space}\label{s5}

\begin{figure}
\centering
\includegraphics[width=11.5cm]{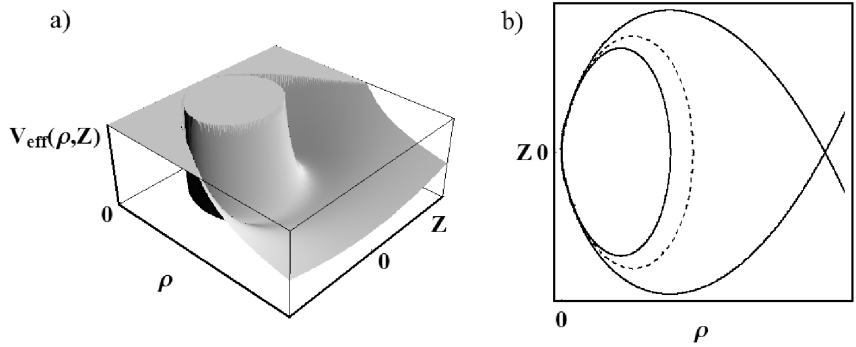}
\caption{a) Graph of the effective potential function (\ref{se5eq1}), with $c_2>0$, for the three dimensional St\"{o}rmer problem for protons. b) Effective potential energy level line (heavy lines), $V_{eff}(\rho, Z)={\bar V}_{eff}(\rho_2)$. 
For particles characterized by the constant of motion $c_2>0$,
the Van Allen inner radiation belt 
is the toric-like surface obtained by rotating the level lines shown in b) around the axis of the dipole field of the Earth. The annular region shown in Figure \ref{fig2}, delimited by the circumferences of radius $\rho_0$ and $\rho_2$, is the $Z=0$ cross 
section of the Van Allen inner radiation belt. The dotted line is the local minimum of the effective potential energy function $V_{eff}(\rho, Z)$.}
\label{fig3}
\end{figure}

The equations of motion of a charged particle in a dipole field --- equations (\ref{eq18}), are derived from
the effective potential function,
\begin{equation}
V_{eff}(\rho, Z)\displaystyle
= {1\over 2 m^2}\left({c_2\over \rho}-m\alpha_1 {\rho\over R^3}\right)^2  
\label{se5eq1}
\end{equation}
where  $R=\sqrt{\rho^2+Z^2}$. 
In Figure \ref{fig3}a), we show the graph of the potential function 
$V_{eff}(\rho, Z)$, and its level lines, for   $c_2>0$.

As we have seen in the previous section, on the equatorial plane 
$Z=0$, the exterior boundary of the trapping region   is the circumference of
radius $\rho_2$, which corresponds to the unique local maximum of the potential
function ${\bar V}_{eff}(\rho)$. The equation  $V_{eff}(\rho, Z)={\bar V}_{eff}(\rho_2)$  defines a bounded region in the $(\rho,Z)$ plane whose closure is a compact set of maximal area\footnote{The compact set is obtained by adding the point $(0,0)$ to the open set defined by $V_{eff}(\rho, Z)={\bar V}_{eff}(\rho_2)$. In the following, compact level sets are always obtained by adding this exceptional point to the open and bounded level sets of the effective potential function $V_{eff}(\rho, Z)$.}, Figure \ref{fig3}b). The compact components of the level sets of the effective Hamiltonian function (\ref{eq20}) are
obtained as the topological product of a compact plane set with the compact sets in the interior 
of the region defined by the equation $V_{eff}(\rho, Z)={\bar V}_{eff}(\rho_2)$. Therefore, any charged particle with an effective energy on a compact component of the effective Hamiltonian is trapped in a torus-like region around the Earth. This trapping region is the Van Allen inner radiation belt of the Earth.

In the rescaled coordinate system introduced in Section \ref{s2}, the level lines $V_{eff}(\rho, Z)={\bar V}_{eff}(\rho_2)$  hit the surface of the Earth for $\sqrt{\rho^2+Z^2}=1$. So, we define the contact points of a specific radiation belt as the points at the surface of the Earth where $V_{eff}(\rho, Z)=0$.
If $c_2>0$, this potential function takes its minimum value ($V_{eff}=0$) along the line $\displaystyle {c_2\over \rho}=m\alpha_1 {\rho\over R^3}$, Figure \ref{fig3}b), and, for $R=1$, we have,   
\begin{equation}
\rho_{c}
= \sqrt{ c_2 \over m\alpha_1}  ={1\over \sqrt{\rho_1}}
\label{se5eq2}
\end{equation}
which corresponds to the latitudes,
\begin{equation}
\theta_c= \arccos \rho_c= \arccos \sqrt{ c_2 \over m\alpha_1}\, .  
\label{se5eq3}
\end{equation}

We now summarize the main features of the dynamics associated with the two degrees of freedom Hamiltonian (\ref{eq20}).

\begin{prop}\label{P2} 
We consider the motion of a nonrelativistic (positively) charged particle in a dipole field. 
If one of the coordinates $Z(0)$ or ${\dot Z}(0)$ of the initial condition is different from zero, and $(\rho(0),Z(0))\not=(0,0)$, then the charged particle has bounded motion, being trapped in a torus-like region around the Earth, provided:
\item{(i)} $H_{eff}(\rho(0),\dot \rho(0),Z(0),\dot Z(0))\le {\bar V}_{eff}(\rho_2)$, $c_2>0$, and $\rho(0)<\rho_2$, where  $\rho_2$ is defined  in (\ref{se4eq6}), the Hamiltonian function $H_{eff}$ is defined in (\ref{eq20}),
and the constant of motion $c_2$ is defined in (\ref{eq19}). 
In the particular case where, $\dot Z(0)=0$, $\dot \rho(0)=0$, 
and $\displaystyle c_2/\rho(0)=m\alpha_1\rho(0)/\sqrt{\rho(0)^2+Z(0)^2}^3$, the
charged particle remains, for every $t\ge 0$, in the plane 
$Z=Z(0)$, and $Z(0)\in [-2m\alpha_1/(3\sqrt{3}c_2),2m\alpha_1/(3\sqrt{3}c_2)]$. If $\dot\phi(0)\not=0$, it rotates around the dipole axis. If $\dot\phi(0)=0$, the particle is at rest in the plane $Z=Z(0)$.
\item{(ii)}
If one of the coordinates $Z(0)$ or ${\dot Z}(0)$ of the initial condition is different from zero, and one of the above conditions is not verified, then $\rho(t)\to \infty$, as
$t \to \infty$, and we have an escape trajectory.
\end{prop}

\begin{proof} If $ Z(0)={\dot Z}(0)=0$, the motion is restricted to the plane perpendicular to the dipole axis, and we obtain the case of Proposition \ref{P1} of the previous section. If $(\rho(0),Z(0))=(0,0)$, the equation of motion (\ref{eq18}) is not defined. In case {\it (i)},  the
initial condition is in a compact component of the effective Hamiltonian, and the motion is bounded for every $t\ge0$. 
If,
$\dot Z(0)=0$, $\dot \rho(0)=0$, 
and $\displaystyle c_2/\rho(0)=m\alpha_1\rho(0)/\sqrt{\rho(0)^2+Z(0)^2}^3$, where $c_2$ is defined in (\ref{eq19}), and $Z(0)\in [-2m\alpha_1/(3\sqrt{3}c_2),2m\alpha_1/(3\sqrt{3}c_2)]$, the differential equation (\ref{eq18}) has one or two fixed points in the plane $Z=Z(0)$.

In case {\it (ii)}, the level sets of the Hamiltonian function are not bounded, and the prolongation of solutions as $t\to \infty$ follows as in case {\it (iv)} of Proposition \ref{P1}.
\end{proof} 
 
\begin{figure}
\centering
\includegraphics[width=10.0cm]{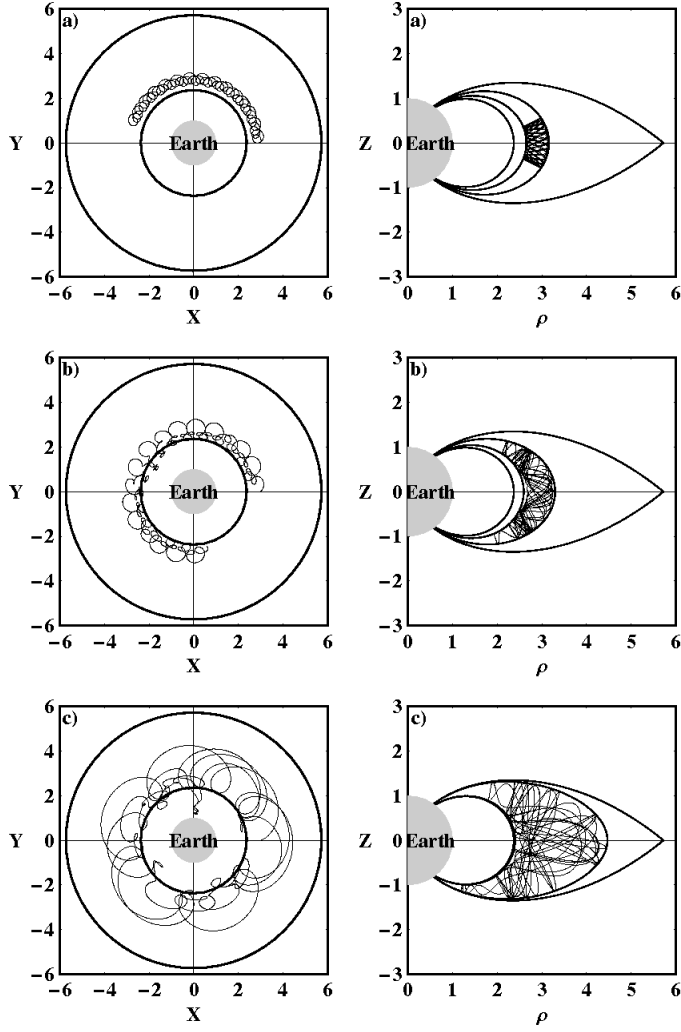}
\caption{Trajectories of three protons  in the Van Allen inner radiation belt, with initial conditions: a) $\rho (0)=3.0$,
$\dot \rho(0)=10.0$, $\phi(0)=0.0$, $\dot \phi (0)=10.0$, $Z(0)=0.5$ and $\dot Z(0)=0.0$; b) $\rho (0)=3.0$,
$\dot \rho(0)=30.0$, $\phi(0)=0.0$, $\dot \phi (0)=10.0$, $Z(0)=0.5$ and $\dot Z(0)=0.0$; c) $\rho (0)=3.0$,
$\dot \rho(0)=80.0$, $\phi(0)=0.0$, $\dot \phi (0)=10.0$, $Z(0)=0.5$ and $\dot Z(0)=0.0$. The radiation belt parameters are $\rho_0=2.370$, $\rho_1=2.861$,
$\rho_2=5.722$, and $c_2=1.776\times 10^{-24}$. 
The total effective energies are: a)
$c_1=500$; b)  $c_1=900$, and,    c)  $c_1=3650$.}
\label{fig4}
\end{figure}

In Figure \ref{fig4}, we show the trajectories of several particles trapped in  a   Van Allen  inner radiation belt characterized by the constant $c_2>0$. The trajectories have been calculate with the 
St\"{o}rmer-Verlet  numerical method (Appendix) with  integration time step $\Delta t=0.00002$, and total integration time $t=2\,$s.  From (\ref{eq19}), it follows that the angular coordinate $\phi (t)$ is given by,
\[
\phi(t)=\displaystyle \phi(0)+\int_0^t \left( {c_2\over m \rho(s)^2}-{\alpha_1\over (\sqrt{\rho(s)^2+Z(t)^2})^3}
\right) dt
\]
and by discretization, we obtain,
\begin{equation}
\phi_{n+1}=\phi_{n}+\Delta t \left( {c_2\over m \rho_n^2}-{\alpha_1\over (\sqrt{\rho_n^2+Z_n^2})^3} \right)  
\label{se5eq5}
\end{equation}
where $\Delta t$ is the discretization time step, $\phi_n=\phi(n \Delta t)$, $\rho_n=\rho(n \Delta t)$,
$Z_n=Z(n \Delta t)$, and $n=0,1,\cdots$. In the trajectories of Figure \ref{fig4}, the angular coordinate $\phi (t)$ has been calculated with (\ref{se5eq5}).

In the plots on the right-hand side of Figure \ref{fig4}, we show  the projection  of the boundary of the level set of the effective Hamiltonian on the $(\rho, Z)$ plane. The trajectories of trapped charged particles are in the interior of these boundary curves.
Particles with low effective energy have trajectories   concentrated  in the vicinity
of the equatorial plane of the Earth, Figure \ref{fig4}a). Increasing the effective energy of the particles, their trajectories  approach the polar
regions of the Earth and eventually hit the surface of the Earth.
By (\ref{se5eq3}), and for the parameter values of Figure \ref{fig4},
the contact points
of the Van Allen inner radiation belt with the Earth are located at the latitudes $\theta_c= \pm 53.76^o$. These simulations suggest that the constant effective energy surface is not filled density by a unique trajectory.

To characterize more precisely  the dynamics of the charged particles trapped in the dipole field of the Earth, we can eventually use KAM techniques, \citep{Lich}, or construct a 
Poincar\'{e} map for the  equations of motion (\ref{eq18}).

 To pursue a KAM approach, \citep{Lich}, the first step is  to find 
an integrable Hamiltonian system leaving invariant a two-dimensional torus
in the four-dimensional phase space of the differential equations  (\ref{eq18}). For that, we develop in Taylor series around the points $(\rho=\rho_1,Z=0)$ and $(\rho=\rho_2,Z=0)$ the second members of the differential equations in (\ref{eq18}). By
Proposition \ref{P1}, at these points the motion is integrable and
periodic in the equatorial plane of the Earth. In the first case, the differential equations  (\ref{eq18}) become,
\begin{equation}
\left\{
\begin{array}{l}\displaystyle
\ddot \rho =-{c_{2}^6\over m^6 \alpha_1^4} (\rho-\rho_1)\\[8pt]\displaystyle
\ddot Z = 0
\end{array}\right.
\label{se5eq6}
\end{equation}
and, in the second case, we obtain,
\begin{equation}
\left\{
\begin{array}{l}\displaystyle
\ddot \rho ={c_{2}^6\over 32 m^6 \alpha_1^4} (\rho-\rho_2)\\[8pt]\displaystyle
\ddot Z = -{3c_{2}^6\over 64 m^6 \alpha_1^4} Z\, .
\end{array}\right.
\label{se5eq7}
\end{equation}
As the solutions of the linear equations (\ref{se5eq6}) and (\ref{se5eq7}) are  unbounded, we loose the property of boundness already contained in Proposition \ref{P2}-{\it (i)} and also the possibility of
having a family of invariant two-dimensional torus in the
four-dimensional phase space of the unperturbed systems (\ref{se5eq6}) and (\ref{se5eq7}). 

The other way of  characterizing the dynamics of the charged particles in the Van Allen inner belt
is to find a  Poincar\'{e}   map for the  equations of motion (\ref{eq18}). 

We consider that  at most one of the coordinates $Z(0)$ and ${\dot Z}(0)$ of the initial condition of the differential equation (\ref{eq18})  is different from zero, and   the particle has bounded motion ($c_2>0$). To avoid degenerate situations, we also consider that, 
$(\rho(0),Z(0))\not=(0,0)$ and the effective energy function is positive, $c_1>0$. Under these conditions,
by Proposition \ref{P2}-{\it (i)}, the initial condition  is on a compact component of the level sets
of the Hamiltonian function (\ref{eq20}). So, the solution of the differential equation (\ref{eq18}) exists, and is defined for all $t\ge 0$ (\cite{chi}, pp. 187; \cite{nemitskii}, pp. 8).

The compact components of the level sets of the Hamiltonian function $H_{eff}$ are  obtained as the topological product of a compact plane set with  the compact level set of the
potential function $V_{eff}(\rho, Z)$. 
These level
sets  are  compact, provided $c_2 >0$. The compact 
level sets of the Hamiltonian function are three-dimensional
compact manifolds embedded in the four dimensional phase space, and
their intersection with the three-dimensional hyperplane $Z=0$
is a two-dimensional compact manifold $\Sigma_{c_{2}}$. This two-dimensional manifold is a compact set in the four dimensional phase space
of the differential equation (\ref{eq18}) and, by (\ref{eq20}), has local coordinates $\rho$ and ${\dot \rho}$. 

In the conditions of Proposition \ref{P2}-{\it (i)}, any orbit initiated in an initial condition on the effective energy
level sets can cross the hyperplane $Z=0$, intersecting transversally the two-dimensional compact manifold $\Sigma_{c_{2}}$. Therefore, the compact two-dimensional manifold  $\Sigma_{c_{2}}$  is a good candidate for a Poincar\'{e} section of the   differential equation (\ref{eq18}). This construction is 
used to justify the existence of two-dimensional conservative Poincar\'{e} maps for two-degrees of freedom Hamiltonian systems, (\cite{Lich}, pp. 17-20). However, there are two main difficulties in proving that $\Sigma_{c_{2}}$ is the domain of a Poincar\'{e} map. The first case, it is
difficult to  prove  that any trajectory that crosses 
transversally the section $\Sigma_{c_{2}}$  will return
to it after a finite time. The second difficult questions concerns the unicity of the trajectories crossing the plane $Z=0$.  

\begin{figure}
\centering
\includegraphics[width=11.0cm]{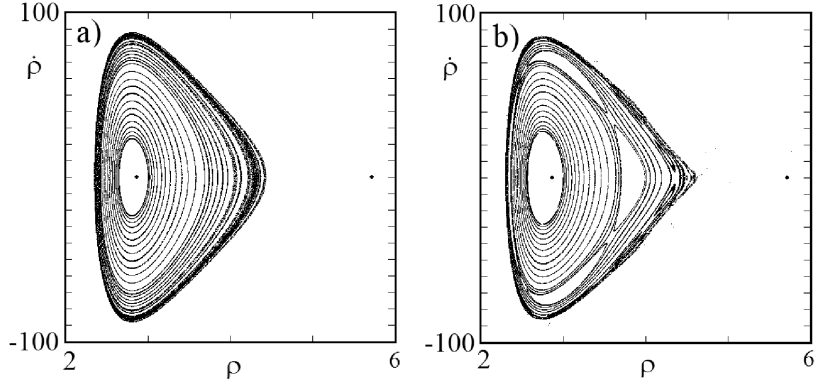}
\caption{Poincar\'{e} maps for the constant  of motion  
$c_2=1.776\times 10^{-24}$.  The Poincar\'{e} map have been calculated with the initial conditions: a)  $\rho (0)=3.0$,
$\phi(0)=0.0$, $\dot \phi (0)=5.47089$, $Z(0)=0.0$ and $\dot Z(0)=0.5$; b)  $\rho (0)=3.0$, $\phi(0)=0.0$, $\dot \phi (0)=10.0$, $Z(0)=0.5$ and $\dot Z(0)=0.5$. In both figures, the initial conditions $\dot \rho(0)$ have been changed in the interval $[0,85]$.
The dots have coordinates  $(\rho_1,0)$ and $(\rho_2,0)$, with $\rho_1<\rho_2$. Both Poincar\'{e} sections have been generated with different
families of initial conditions. This suggests that there are different particle trajectories with the same effective energy that cross the two-dimensional section $\Sigma_{c_{2}}$ at the same point.
}
\label{fig5}
\end{figure}

From Proposition \ref{P2}-{\it (i)}, it follows that there are orbits of trapped particles that never cross $\Sigma_{c_{2}}$. However, due to the
invariance of  the differential equation (\ref{eq18}) for
transformations of $t$ into $-t$, any initial condition on the plane
$Z=0$ with ${\dot Z}\not=0$, has  a prolongation for positive and 
negative values of $t$, crossing  transversally the plane $Z=0$ at most once.
With this property, we can test numerically the structure of the orbits that cross the set $\Sigma_{c_{2}}$.
In Figure \ref{fig5}, we show the computed orbits on the Poincar\'{e}  section $\Sigma_{c_{2}}$, for two families of initial conditions and the same value of the constant of motion  $c_2$.

In Figure \ref{fig5}a), we have chosen initial conditions on the plane $Z(0)=0$, with 
${\dot Z}(0)\not =0$. For all the analysed initial conditions, the orbits crossed several times the surface of section $\Sigma_{c_{2}}$. Due to the structure of orbits, the motion appears to be quasi-periodic. However, if the initial conditions are away from the plane $Z(0)=0$ and for the same value of $c_2$, the topology of the orbits
on the section $\Sigma_{c_{2}}$ changes drastically, Figure \ref{fig5}b). Comparing Figure \ref{fig5}a) with \ref{fig5}b), the structure of the orbits are incompatible, in the sense that different orbits cross $\Sigma_{c_{2}}$ at the same point. This shows that  $\Sigma_{c_{2}}$
can not be the domain of a Poincar\'{e}  map for the differential equations
(\ref{eq18}). On the other hand, the structure of orbits in Figure \ref{fig5}b)
suggests the existence of transversal homoclinic intersections and therefore  chaotic motion inside the trapping region of 
the St\"{o}rmer problem.

\section{Conclusions}\label{s6}

The  Hamiltonian dynamical system describing the 
motion of a charged particle in a dipole field (St\"{o}rmer problem) has been reduced 
to a two-degrees of freedom system. This reduced system has two constants of motion. It has been shown that the trajectories of charged particles can be periodic and quasi-periodic, and that, for
a suitable choice of the initial conditions in between a minimal and maximal height from the equatorial plane, the planes perpendicular to the dipole axis are invariant for the motion of charged particles. 

From a more global point of view, the trajectories of the charged particles in the Earth dipole field can be trapped in a torus-like region surrounding the Earth,
or can be scattered and escape to infinity. The  torus-like trapping region
around the Earth can be interpreted as the Van Allen inner radiation belt, as measured by particle detectors in spacecrafts. The physical effects associated with radiation phenomena intrinsic to accelerated charged particles ({\it Bremsstrahlung}) account for the phenomena of radiation aurorae. The numerically computed trajectories of trapped charged particles suggests the existence of chaotic motion inside the Van Allen inner belts around the dipole axis of the Earth. The properties of the St\"{o}rmer dynamical systems are
summarized in  Propositions \ref{P1} and \ref{P2}.

From the applied physics point of view, the St\"{o}rmer problem deviates from
the real situation in essentially three ways. In the first case, the magnetic field of the Earth has a strong quadrupolar component that was not considered. The second drastic simplification has to do with
the fact that the dipole axis is not coincident with the rotation axis of the Earth. The rotation of the Earth introduces a time periodic forcing into the equations of motion. 
In the third simplification, we have considered implicitly that  charged particles do not radiate ({\it Bremsstrahlung}) when subject to accelerating forces. However,   the three main effects observed in the Earth magnetosphere, radiation belts, radiation aurorae and South Atlantic anomaly,  
are described by the simplified model. 

The existence of periodic, quasi-periodic and chaotic trajectories 
in the  St\"{o}r\-mer problem together with the {\it Bremsstrahlung}
effect, shows that radiation belts make a shield protection for the high energy charged particles arriving at Earth. These particles loose kinetic energy by {\it Bremsstrahlung} when they are scattered and when they are trapped in the Van Allen inner radiation belts. 

From the mathematical point of view, the St\"{o}rmer problem
poses some open problems for the dynamics of two-degrees of freedom Hamiltonian systems. The St\"{o}rmer dynamical systems is not generated from the perturbation of a two-dimensional torus in a four-dimensional phase space, suggesting the existence of non KAM mechanism for the generation of bounded motion in two-degrees of freedom Hamiltonian systems. On the other hand, the arguments used in  the construction of Poincar\'{e} sections in  two-degrees of freedom Hamiltonian systems with two conservation laws lead to
the non-unicity of the orbits on these surfaces of section.  

{}


\section*{Appendix}\label{ap}

To integrate numerically the equations of 
motion (\ref{eq18}) and (\ref{se4eq1}), we have used the explicit St\"{o}rmer-Verlet method of order 2,
(\cite{Gni}, pp. 14 and 177),
\[
\left\{
\begin{array}{l}\displaystyle
p_{n+1/2}=p_{n}-\Delta t {\partial H\over \partial q}(q_i)\\[8pt]\displaystyle
q_{n+1}=q_{n}+\Delta t {\partial H\over \partial p}(p_{n+1/2})\\[8pt]\displaystyle
p_{n+1}=p_{n+1/2}-\Delta t {\partial H\over \partial q}(q_{i+1}) 
\end{array}\right.
\label{se3eq1}
\]
where $H(q_1,\ldots,q_n,p_1,\ldots,p_n)$ is the Hamiltonian function, and $\Delta t $ is the integration time step. For example, for equation (\ref{se4eq1}), the St\"{o}rmer-Verlet method reduces to,
\[
\left\{
\begin{array}{l}\displaystyle
\rho_{n+1}=\rho_{n}+\Delta t \dot \rho_{n}+ {\Delta t^2\over 2}f(\rho_{n})\\[8pt]\displaystyle
\dot \rho_{n+1}=\dot \rho_{n}+{\Delta t \over 2}f(\rho_{n})+ {\Delta t \over 2}f(\rho_{n+1})\\[5pt]
\end{array}\right.
\]
where $\rho_n=\rho(n\Delta t)$, $\dot \rho_n=\dot \rho(n\Delta t)$, $f(\rho)= \displaystyle -{dV_{eff}\over d\rho}$. This method is symplectic, being area preserving. 

The St\"{o}rmer-Verlet method has the advantage of being explicit, and the integration accuracy is obtained by decreasing the time step $\Delta t$. However, it does not conserve the energy function. We have tested other higher order numerical integrators but, with this
St\"{o}rmer-Verlet method, the 
overall behaviour of the solutions is closer to the exact results.  
For a detailed theoretical discussion about the St\"{o}rmer-Verlet method see \cite{Tupper}.

\subsection*{Acknowledgment}
This work has been partially supported by the POCTI Project /FIS/10117/2001 (Portugal) and by a pluriannual funding grant to 
GDNL.

\end{document}